\documentclass[aps,pra,reprint,showpacs,groupedaddress]{revtex4-1}
\usepackage{graphicx}
\usepackage{amsmath,amssymb}
\begin{document}

\title{A simplified 461-nm laser system using blue laser diodes and a hollow cathode lamp for laser cooling of Sr}
\author{Yosuke Shimada}
\author{Yuko Chida}
\author{Nozomi Ohtsubo}
\author{Takatoshi Aoki}
\author{Yoshio Torii}
\email[]{ytorii@phys.c.u-tokyo.ac.jp}
\affiliation{Institute of Physics, The University of Tokyo, 3-8-1 Komaba, Meguro-ku, Tokyo 153-8902, Japan}
\date{\today}

\begin{abstract}
We develop a simplified light source at 461 nm for laser cooling of Sr without frequency-doubling crystals but with blue laser diodes. An anti-reflection coated blue laser diode in an external cavity (Littrow) configuration provides an output power of 40 mW at 461 nm. Another blue laser diode is used to amplify the laser power up to 110 mW by injection locking. For frequency stabilization, we demonstrate modulation-free polarization spectroscopy of Sr in a hollow cathode lamp. The simplification of the laser system achieved in this work is of great importance for the construction of transportable optical lattice clocks.

\end{abstract}
\maketitle

\section{Introduction}
The range of applications with laser-cooled strontium has tremendously expanded during the past decade. The Sr optical lattice clock \cite{Takamoto}, where an extremely narrow ($\sim$mHz) intercombination line is used as a clock transition, is now regarded as a candidate for the future redefinition of the second. The techniques to create quantum degenerate samples of Sr have established for all the stable isotopes ($^{84}$Sr \cite{84Sr1, 84Sr2}, $^{86}$Sr \cite{86Sr}, $^{87}$Sr \cite{87Sr}, and $^{88}$Sr \cite{88Sr}). Recently, the creation of ultracold Sr$_{2}$ molecules in the electronic ground state was realized \cite{Sr2}. Precise measurement of gravity using Bloch oscillations of Sr atoms in an optical lattice \cite{Poli} benefited from the very small s-wave scattering length of $^{88}$Sr \cite{Martinez}, which is ideal for atom optics experiments.

So far, 461-nm light for laser cooling of Sr on the $(5s^{2}) ^{1}S_{0} \to (5s5p) {}^{1}P_{1}$ transition has been produced by second-harmonic generation (SHG) using doubling crystals such as KNbO$_{3}$ \cite{Polzik}, BIBO \cite{Cruz}, and PPKTP \cite{LeTargat}.  Single-pass PPLN waveguides, which deliver about 80 mW of 461-nm light with a fundamental power of 250 mW, have also been developed \cite{Akamatsu}. 

Recently, blue laser diodes operating in the range of 450$\sim$460 nm are commercially available. In this paper we report the construction of a simple 461 nm light source based on blue laser diodes. We demonstrate that a blue laser diode can, with the help of injection locking, deliver up to 110 mW of 461 nm light, which is sufficient for Zeeman slowing and magneto-optical trapping of Sr \cite{Courtillot}.  For spectroscopy of Sr we adopt a hollow cathode lamp, which provides a Sr vapor with sufficiently large optical density. We demonstrate, for the first time to our best knowledge, polarization spectroscopy of Sr in a hollow cathode lamp as a modulation-free and hence inexpensive way of frequency stabilization. The obtained  dispersion (error) signal, contrary to that obtained by a standard wavelength-modulation technique, possesses a long tail, which is advantageous for robust frequency locking. The simplification of the laser system at 461 nm achieved in this work would facilitate the Sr laser cooling experiments and is of crucial importance especially for the construction of transportable optical lattice clocks.

\section{External cavity laser diode at 461 nm}
\begin{figure*}[tb]
\begin{center}
\includegraphics[width=155mm]{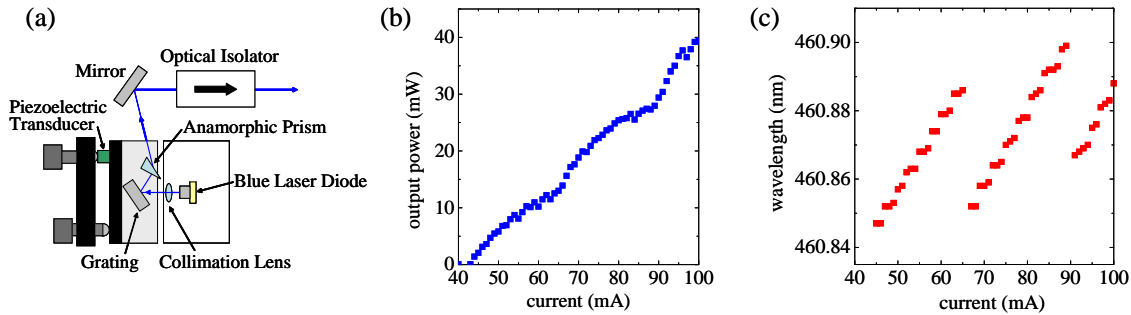}
\end{center}
\caption{(a) Schematic of the external cavity laser diode using a blue laser diode.  (b) Power and (c) wavelength characteristics of the external cavity laser diode tuned around 460.862 nm at 25 $^\circ$C}
\label{Fig1}
\end{figure*}

For laser cooling experiments, external cavity laser diodes (ECLDs) have widely been used  since they provide improved frequency tunability and reduced intrinsic linewidths \cite{Ricci}. Figure \ref{Fig1}(a) shows a schematic of the ECLD we constructed using a blue laser diode (LD) with a peak wavelength of 457 nm (test sample from Nichia Corporation).
The blue LD is anti-reflection (AR) coated and set in a Littrow configuration using a collimation lens with a focal length of 4.0 mm (Thorlabs C671TME-A) and a holographic grating of 2400 grooves/mm (Thorlabs GH13-24V). The polarization of the light from the LD is parallel to the grooves of the grating and the diffraction efficiencies of the 0th order (output coupling) and the 1st order (optical feedback) are 38\% and 46\%, respectively \cite{Note1}. The collimated laser beam has a diameter of about 2 mm (1 mm) in the horizontal (vertical) direction, and the grating resolution is $\Delta \lambda=\lambda/N\sim0.08$ nm, where $N\sim6000$ is the number of grooves illuminated by the laser beam. The alignment of the grating is accomplished by a commercial mirror mount, which is fixed to the laser mount through four stainless steel rods (not shown in the figure). A piezoelectric transducer (PZT) is sandwiched between the front plate of the mirror mount and the screw which is used for coarse wavelength tuning. An anamorphic prism transforms the elliptical beam from the LD into a nearly circular beam.

The power and wavelength characteristics of the ECLD tuned around the Sr $(5s^{2}) ^{1}S_{0} \to (5s5p) {}^{1}P_{1}$ transition (460.862 nm) at a temperature of 25 $^\circ$C are shown in Fig.\ref{Fig1}(a) and (b), respectively. The threshold current is 43 mA and the output power reaches 40 mW at 100 mA diode current. Without grating feedback we observed no lasing with currents up to 140 mA. We checked with a Fabre-Perot resonator that the ECLD operates in single mode except in the vicinity of mode hopping. As expected the range of the lasing wavelength ($\sim$0.05 nm) is limited to within the grating resolution of 0.08 nm. 

The dependence of the laser wavelength on the diode current is explained as follows. The increase in the diode current generally increases the refractive index of the diode chip and hence increases the optical length of the internal (diode-chip) cavity. If we assume that the increase $\Delta L$ of the optical length is proportional to the diode current $I$ (i.e., $\Delta L=\alpha I$, where $\alpha$ is a constant), the frequency shift $\Delta \nu$ of the cavity modes is given by 
\begin{equation}
\Delta \nu =-\nu \frac{\Delta L}{L}=-\beta I,
\end{equation}
where $\nu$ is the laser frequency, $L$ is the length of the internal or external cavity, and $\beta=\nu \alpha/L$ represents the ratio of the mode frequency shift to the diode current. In the case of our ECLD, the optical length of the external cavity ($\sim$20 mm) is about seven times longer than that of the internal cavity ($\sim$2.8 mm). Therefore, $\beta$ for the internal cavity is about seven times larger than that for the external cavity. Even thought the laser diode is AR coated, the mode selection due to the internal cavity still works and determines in which external cavity mode the ECLD operates. As a result, mode hops occur between adjacent external cavity modes separated by the free spectral range (FSR) of 0.005 nm (7 GHz) when we sweep the diode current. The large mode hops at 66 mA and 90 mA, where kinks in the output power are seen, are the mode hops between external cavity modes separated by 0.04 nm (50 GHz), which is the FSR of the internal cavity. The values of $\beta$ obtained from Fig.\ref{Fig1}(c) and a separate measurement using a Fabre-Perot resonator are 3 GHz/mA and 0.4 GHz/mA for the internal and external cavity, respectively.

Frequency scanning of the ECLD is accomplished by applying a ramp voltage to the PZT. The mode-hop-free scanning range (MHFSR) was measured to be about 8 GHz, which is close to the FSR of the external cavity. Much larger MHFSRs would be achieved if we simultaneously sweep the diode current such that the shift of the internal cavity modes coincides with the shift of the external cavity modes. We did, however, no such current sweep in the experiments described below because a 8-GHz MHFSR is large enough compared with the Doppler width ($\sim$2 GHz) and the isotope shifts (-270.8 MHz for $^{84}$Sr and -124.5 MHz for $^{86}$Sr with respect to $^{88}$Sr) of the ${}^{1}S_{0} \to {}^{1}P_{1}$ transition.  

\section{Power amplification by injection locking}
\begin{figure*}[tbh]
\begin{center}
\includegraphics[width=150mm]{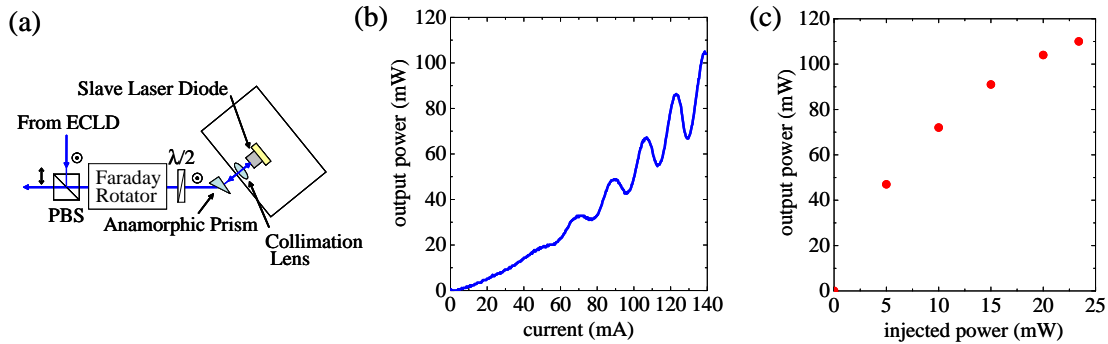}
\end{center}
\caption{(a) Setup for injection locking of the slave laser diode. PBS, polarizing beam splitter; $\lambda$/2, halfwave plate. (b)Dependence of the slave laser power on the laser current for 20 mW injection power. (c)Dependence of the slave laser power on the injected (master) laser power for 140 mA slave laser current.}
\label{Fig2}
\end{figure*}

The output power of the ECLD is not sufficient for laser cooling of Sr as it is, so we employ the injection locking technique \cite{Kobayashi} using another AR-coated blue LD as a slave laser \cite{Note2}. As shown in Fig.\ref{Fig2}(a), part of the light from the ECLD is injected into the slave laser by using a polarizing beam splitter, a Faraday rotator, and a halfwave plate. An anamorphic prism is placed in front of the slave laser for mode matching. 

The output power of the slave laser as a function of the slave laser current is shown in Fig.\ref{Fig2}(b) for 20 mW injection power. The output power reaches 100 mW at around 140 mA slave laser current. Above the slave laser current of 40 mA, the output power oscillates as a function of the injected power. Despite this oscillatory behavior the slave laser frequency is always locked to the master laser regardless of the slave laser current (we checked this not only by monitoring the spectrum of the slave laser light using a Fabre-Perot resonator but also by observing the absorption spectra of Sr, as described below, both with the master laser and with the slave laser). This means that the slave laser acts rather as a master-oscillator power amplifier (MOPA) like a tapered amplifier \cite{Wilson}. The period of the oscillation in the output power seen in Fig.\ref{Fig2}(b) is about 20 mA, which corresponds to the shift of the internal cavity modes by one FSR (50 GHz). This indicates that the efficiency of the power amplification is maximized when one of the frequencies of the internal cavity modes matches the frequency of the master laser.

Figure \ref{Fig2}(c) shows the dependence of the slave laser power on the injected laser power for a slave laser current of 140 mA. Actually, the slave laser current was slightly adjusted around 140 mA for each injection power such that the output power was maximized. This is because the frequencies of the internal cavity modes shift depending on the intensity of the laser light inside the diode chip. The gain of power amplification is about 10 at 5 mW injection power, but the slave laser power saturates at around 110 mW above the injection power of $\sim$20 mW. It would be preferable to prepare another slave laser if one needs a laser power more than 100 mW. For example, if 30 mW of the master laser power is available for injection locking, two of the slave lasers would give a total output power of 180 mW with an injection power of 15 mW for each slave laser. 

\section{Sr hollow cathode lamp}
The vapor pressure of Sr at room temperature is quite low, therefore heatpipe ovens \cite{Li} or atomic beams from ovens \cite{Courtillot} or dispensers \cite{Bridge} have widely been used for spectroscopy of Sr. Alternatively, a hollow cathode lamp (HCL) offers a simple way of obtaining a Sr vapor suitable for spectroscopy \cite{Aoki}.

\begin{figure}[tbbb]
\begin{center}
\includegraphics[width=70mm]{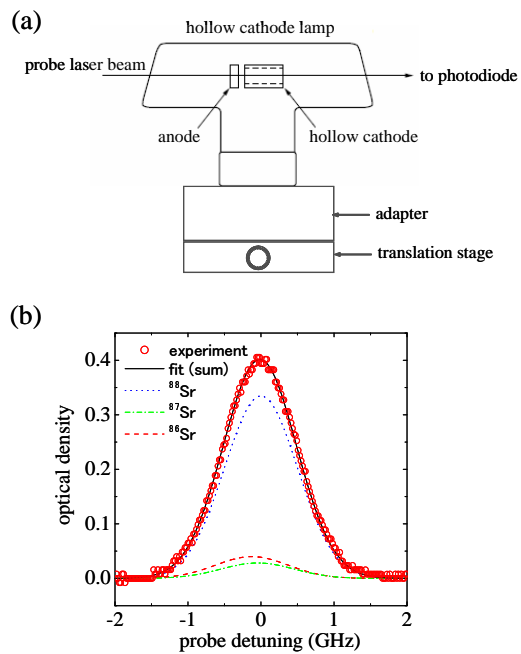}
\end{center}
\caption{(a) Setup for measuring the spatial distribution of Sr atoms in the hollow cathode. (b) Typical Doppler-broadened absorption spectrum of the Sr ${}^{1}S_{0} \to {}^{1}P_{1}$ transition measured in the center of the hollow cathode. The solid line is a fit considering the three major isotopes of $^{88}$Sr (82.58\%), $^{87}$Sr (7.00\%), and $^{86}$Sr(9.86\%).}
\label{Fig3}
\end{figure}

We adopt a commercial see-through HCL (Hamamatsu L2783-38Ne-Sr) as a frequency reference of the Sr $(5s^{2}) ^{1}S_{0} \to (5s5p) {}^{1}P_{1}$ transition (natural linewidth: 32 MHz). The HCL is filled with a Ne buffer gas of 5-10 Torr. The anode is ring-shaped, and the cathode, which contains Sr, has a length of about 20 mm and a bore diameter of 3 mm [see Fig.\ref{Fig3}(a)]. We need to apply a voltage of 190 V across the anode and the cathode for a discharge to start, while the discharge can be sustained at a voltage of about 160 V for discharge currents up to 20 mA. Positive neon ions created in the discharge strike the inner surface of the cathode and sputter Sr atoms. 
 
Unlike the vapor cells, the spatial distribution of the sample atoms (or ions) is not uniform inside the hollow cathode \cite{Gerstenberger}. To determine the spatial distribution of neutral Sr atoms in the ground ($(5s^{2}){}^{1}S_{0}$) state inside the hollow cathode, we performed absorption spectroscopy on the $(5s^{2}){}^{1}S_{0} \to (5s5p){}^{1}P_{1}$ transition. A sketch of the setup is shown in Fig.\ref{Fig3}(a).  A probe laser beam, which was derived from the ECLD we described above, was spatially filtered using a single mode fiber, sent through the hollow cathode, and detected by a photodiode. At the position of the hollow cathode, the probe beam had a $1/e^{2}$ waist of 0.13 mm and an intensity of about 20 mW/cm$^{2}$, which is below the saturation intensity of 42 mW/cm$^{2}$ \cite{Note3}. The HCL was mounted on a translation stage which moves in the horizontal direction perpendicular to the probe beam.

Figure \ref{Fig3}(b) shows a typical Doppler-broadened absorption spectrum measured in the center of the hollow cathode with a discharge current of 20 mA. The vertical axis in the figure represents the optical density (OD) defined as $\text{OD}=-\ln(P/P_{0})$, where $P$ and $P_{0}$ are the transmitted probe beam powers with the HCL on and off, respectively. From a fit assuming Gaussian (Maxwell-Boltzmann) distributions for the three major isotopes ($^{88}$Sr, $^{87}$Sr, and $^{86}$Sr), the temperature of the Sr atoms inside the hollow cathode is estimated to be 530 K. We observed no discernible change in the width of the absorption spectrum when we changed the distance of the probe beam from the cathode center, indicating that the temperature distribution of the Ne buffer gas is almost uniform inside the hollow cathode. 

\begin{figure}[tbh]
\begin{center}
\includegraphics[width=85mm]{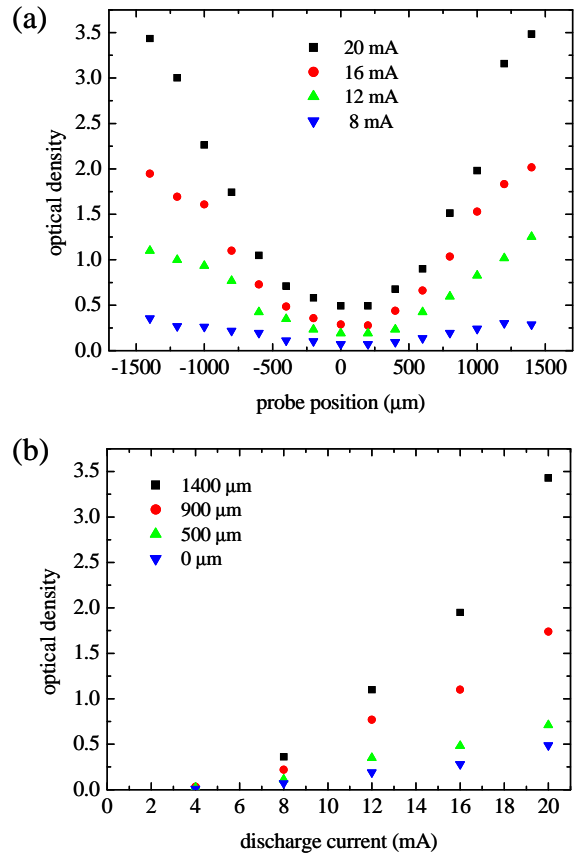}
\end{center}
\caption {(a) Spatial variations of the peak optical density of Sr in the hollow cathode for various discharge currents. (b)Dependence of the peak optical density of Sr on the discharge current at distances of 0, 500, 900, and 1400 $\mu$m from the center of the cathode.}
\label{Fig4}
\end{figure}

The spatial variations of the OD for various discharge currents are shown in Fig.\ref{Fig4}(a). The OD significantly depends on the distance from the center of the cathode: at 1400 $\mu$m from the center (just near the wall of the cathode) it is about seven times larger than that in the center for 20 mA discharge current. This strong spatial variation of the OD presumably comes from the extremely slow diffusion of the sputtered Sr atoms; the estimated mean free path of Sr atoms at 5-10 Torr of the Ne buffer gas is of the order of 10 $\mu$m, which is much shorter than the diameter of the bore (3 mm), so the sputtered Sr atoms spend most of their time in the vicinity of the cathode wall \cite{Baguer2}.

The dependence of the OD on the discharge current exhibits nonlinear behavior as shown in Fig.\ref{Fig4}(b). A possible origin of this nonlinearity is the increase in the sputtering yield (the number of sputtered atoms per incident particle) with increasing discharge current \cite{Baguer2}: as we increase the discharge current, the distributions of the positive neon ions and electrons change such that the electric field near the cathode wall becomes stronger \cite{Baguer}, resulting in the increase in the average kinetic energy of the incident positive neon ions and thereby in the increase in the sputtering yield.

\begin{figure*}[bbbt]
\begin{center}
\includegraphics[width=155mm]{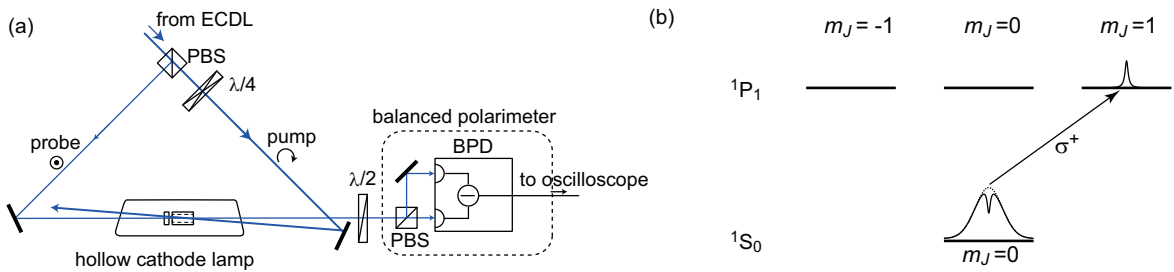}
\end{center}
\caption{(a) Setup for polarization spectroscopy of Sr in a hollow cathode lamp. The balanced polarimeter consists of a balanced photodetector (BPD) and a polarizing beamsplitter (PBS). (b) Relevant energy diagram of Sr for the bosonic isotopes (with no hyperfine structure). The effect of the $\sigma^{+}$ pump beam on the velocity distribution for each magnetic sublevel is illustrated.}
\label{Fig5}
\end{figure*}

\section{Polarization spectroscopy of Sr}
Although frequency-modulation (FM) spectroscopy \cite{Bjorklund} or wavelength-modulation (WM) spectroscopy \cite{Supplee} is a straightforward way of obtaining an error signal for laser frequency stabilization, we choose instead modulation-free polarization spectroscopy \cite{Wieman}, which also provides a dispersive error signal, to make the whole laser system (including the electronics) even simpler. We note that polarization spectroscopy has the additional merit that it offers a wide feedback bandwidth, enabling efficient laser linewidth reduction \cite{Torii}.   

Polarization spectroscoy of Sr was firsr demonstrated on the 689-nm intercombination [$(5s^{2}){}^{1}S_{0} \to (5s5p){}^{3}P_{1}$] transition in a hollow cathode discharge \cite{Tino}. As for the 461-nm [$(5s^{2}) ^{1}S_{0} \to (5s5p) {}^{1}P_{1}$] transition, the only report we are aware of is Javaux {\it at al.} \cite{Javaux}, where polarization spectroscoy is performed not in a hollow cathode discharge but in a vapor cell equipped with a dispenser \cite{Bridge}. 

We performed polarization spectroscopy of Sr on the $(5s^{2}) ^{1}S_{0} \to (5s5p) {}^{1}P_{1}$ transition in a hollow cathode discharge with the setup illustrated in Fig.\ref{Fig5} (a). A circularly-polarized (${{\sigma }^{+}}$) pump beam and a linearly-polarized probe beam intersect with each other at an angle of 0.01 rad in the hollow cathode (we choose the quantization axis along the propagation direction of the probe beam).  As depicted in Fig.\ref{Fig5} (b), the ${{\sigma }^{+}}$ pump beam creates a hole and a bump in the velocity distributions of the atoms in the $^{1}S_{0}, m_{J}=0$ and $^{1}P_{1}, m_{J}=+1$ state, respectively (we assume here that the Sr atoms are bosons, such as $^{88}$Sr, and have no nuclear spin and hence no hyperfine structure). As a result, the probe beam, while passing through the hollow cathode, experiences circular birefringence and the polarization axis of the probe beam rotates. This rotation of the polarization axis is monitored by a balanced polarimeter composed of a polarizing beamsplitter and a balanced (differential) photodetector. A halfwave plate is placed in front of the balanced polarimeter and its angle is adjusted such that the balanced polarimeter gives no signal without the pump beam \cite{Yoshikawa}.

On the assumption that the effect of velocity-changing collisions (VCCs) \cite{Smith} is negligible, the signal obtained by the balanced polarimeter is derived from the Kramers-Kronig dispersion relation as (see Appendix)
\begin{equation} 
\Delta I=-2\Delta {I}_{\max}\frac{x}{1+{{x}^{2}}},
\label{Eq2}
\end{equation}
where $x=2(\omega -{{\omega }_{0}})/\Gamma$ is a normalized probe detuning from resonant frequency ${{\omega }_{0}}$, $\Gamma $ is the (power- and collision-broadened) homogeneous linewidth (FWHM) of the transition, and 
\begin{equation}
\Delta {{I}_{\max }}=\frac{{I}_{T}}{6}\Delta \text{OD}_{\max}
\label{Eq3}
\end{equation}
is the maximum amplitude of the signal, where  ${{I}_{T}}$ is the signal corresponding to the transmitted probe laser power and $\Delta \text{OD}_{\max}$ is the change of the peak OD due to the pump beam (i.e., the depth of the Lamb-dip in terms of the OD).

\begin{figure}[tbb]
\begin{center}
\includegraphics[width=65mm]{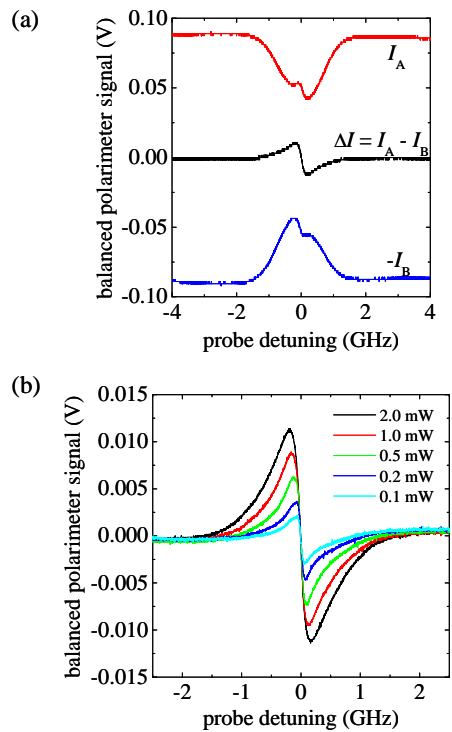}
\end{center}
\caption{(a)Typical signals obtained by the balanced polarimeter. The signals labeled $I_{A}$ and $-I_{B}$ are obtained when one of the balanced photodetector inputs is blocked. The signal labeled $\Delta I=I_{A}-I_{B}$ corresponds to a polarization spectrum expressed by Eq.(\ref{Eq2}). (b)Polarization spectra for various pump beam powers.}
\label{Fig6}
\end{figure}

\begin{figure}[hbt]
\begin{center}
\includegraphics[width=80mm]{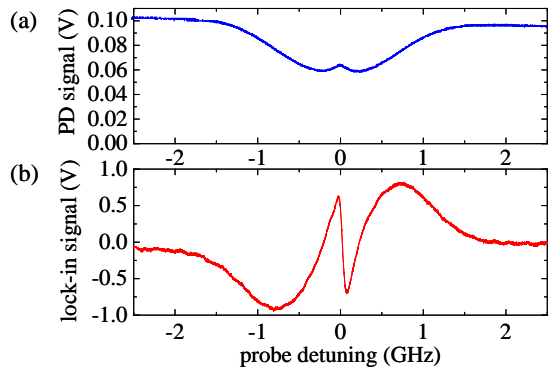}
\end{center}
\caption{ (a) Saturated-absorption spectrum on the Sr ${}^{1}S_{0} \to {}^{1}P_{1}$ transition and (b) its derivative obtained by wavelength-modulation spectroscopy. The time constant of the lock-in amplifier was 10 ms and the sweep rate of the laser frequency was 500 MHz/s}
\label{Fig7}
\end{figure}

Figure \ref{Fig6}(a) shows typical signals obtained by the balanced polarimeter with pump and probe beam powers (intensities) of 2.0 mW (16 W/cm$^{2}$) and 0.15 mW (1 W/cm$^{2}$), respectively. The $1/e^2$ radius of both the pump and probe beam was 90 $\mu$m and the position of the HCL was adjusted so that the OD for the probe beam became $\sim$1. The signals labeled $I_{A}$ and $-I_{B}$ in Fig.\ref{Fig6}(a) were obtained with one of the balanced photodetector inputs being blocked, while the dispersive signal labeled $\Delta I=I_{A}-I_{B}$, which corresponds to a polarization spectrum expressed by Eq.(\ref{Eq2}), was obtained with both of the balanced photodetector inputs unblocked. The signal corresponding to the transmitted probe power is given by $I_{T}=I_{A}+I_{B}$, which is needed to compare the observed spectrum with Eq.(\ref{Eq3}). 

Here, we compare the maximum amplitude of the observed polarization spectrum in Fig.\ref{Fig6}(a) with the value calculated by Eq.(\ref{Eq3}) quantitatively. From the observed depth of the Lamb-dip ($\Delta \text{OD}_{\max}\sim0.4$) and the signal corresponging to the transmitted probe power ($I_{T}\sim0.1$ V), the maximum amplitude $\Delta {I}_{\max}$ is calculated from Eq.(\ref{Eq3}) to be 0.007 V, whereas the observed value of $\Delta {{I}_{\max }}$ in Fig.\ref{Fig6}(a) is 0.012 V. This discrepancy is presumably due to the presence of VCCs and/or the fact that the probe beam intensity is larger than the saturation intensity and the Kramers-Kronig dispersion relation cannot be strictly applied.  

Figure \ref{Fig6}(b) shows the polarization spectra for various pump beam powers with the probe beam power kept at 0.15 mW. Both the width and amplitude of the dispersion curve increase with increasing pump beam power. For weak pump beam powers below 0.2 mW, we can notice that the dispersion curve becomes asymmetric due to the contributions from $^{86}$Sr and $^{87}$Sr, as has been reported in Ref.\cite{Javaux}.  The quantitative explanation of the dependences of the width and the height of the dispersion curve on the pump beam power requires detailed knowledge on the collisional properties of Sr and Ne, and is beyond the scope of this paper. A detailed study on polarization spectroscopy of Sr in the hollow cathode discharge will be reported elsewhere.

We also demonstrated Doppler-free WM spectroscopy of Sr using a HCL. The wavelength of the 461-nm light from the ECLD can be modulated just by modulating the current of the laser diode. This makes a marked contrast with the case of SHG using an enhancement cavity \cite{Polzik, Cruz, LeTargat}, where the fundamental light cannot not be wavelength modulated for the cavity to keep resonance with the fundamental light. 

For Doppler-free WM spectroscopy we used the same setup for polarization spectroscopy [Fig.\ref{Fig5}(a)] with the modification that the balanced polarimeter was replaced with a single photodetector. The powers (intensities) of the pump and probe beam were 3.0 mW (24 W/cm$^{2}$) and 25 $\mu$W (200 mW/cm$^{2}$), respectively. The laser frequency was modulated at 7 kHz with an amplitude of 25 MHz by modulating the laser diode current. The resulting modulation in the probe beam power was detected by the photodetector and fed to a lock-in amplifier. Figure \ref{Fig7} (a) and (b) show a saturated-absorption (SA) spectrum and its derivative (WM spectrum) obtained by the lock-in amplifier, respectively. We can notice that a sharp dispersion curve, whose width is determined by the homogeneous width, is embedded in a broad dispersion curve originating from the Doppler profile of the SA spectrum. 

The suitability of polarization spectroscopy for laser frequency stabilization is apparent if we compare the signal obtained by polarization spectroscopy [Fig.\ref{Fig6}(b)] with that obtained by WM spectroscopy [Fig.\ref{Fig7}(b)]. The latter has two extra zero-crossing points close to the resonance frequency, which narrow the locking range of the laser. The former, on the other hand, has no such extra zero-crossing point and is ideal for robust laser frequency stabilization.

\section{conclusion}
We have developed a simple and compact laser system at 461 nm which does not rely on SHG.  An external cavity laser diode using an AR-coated blue laser diode, which has recently been commercially available, delivers an output power of 40 mW. Injection locking allows us to amplify the power up to 110 mW, which is sufficient for laser cooling of Sr on the $(5s^{2}) ^{1}S_{0} \to (5s5p) {}^{1}P_{1}$ transition. We adopted a hollow cathode lamp for spectroscopy of Sr and observed a strong spatial variation of the optical density of Sr atoms inside the hollow cathode, which is of obvious importance in performing spectroscopy of Sr with good signal-to-noise ratios. We performed modulation-free polarization spectroscopy of Sr in a hollow cathode lamp, obtaining a dispersion signal suitable for a robust frequency locking. The simplification of the laser system at 461 nm achieved in this work would greatly facilitate various Sr laser cooling experiments including the construction of a portable optical lattice clock.

\appendix*
\section{polarization signal obtained by the balanced polarimeter}
In Doppler-free polarization spectroscopy of bosonic Sr (with no hyperfine structure), the ${{\sigma }^{+}}$ pump beam creates a hole and a bump in the velocity distribution of atoms in the $^{1}S_{0}, m_{J}=0$ and $^{1}P_{1}, m_{J}=+1$ state, respectively [Fig.\ref{Fig5} (b)]. As a result, the ${{\sigma }^{+}}$ and ${{\sigma }^{-}}$ components of the probe beam experience different absorption coefficients, ${{\alpha }^{+}}$ and ${{\alpha }^{-}}$, and different refractive indices, ${{n}^{+}}$ and ${{n}^{-}}$, while passing through the hollow cathode. The difference of the refractive index $\Delta n={{n}^{+}}-{{n}^{-}}$ is related to the rotation angle $\Delta \theta $ of the probe-beam polarization axis by $\Delta \theta =\Delta n{{k}_{0}}l/2$, where ${{k}_{0}}$ is the wave number of the probe beam in vacuum and $l$ is the length of the hollow cathode (we assume that the density profile of Sr atoms in the hollow cathode is uniform along the axial direction). The rotation angle $\Delta \theta$ is converted to the electronic signal $\Delta I$ by the balanced polarimeter with the relation \cite{Yoshikawa}
\begin{equation}
\Delta I=2{{I}_{0}}\exp (-\bar{\alpha }l)\Delta \theta,
\end{equation}
where ${{I}_{0}}$ is the signal corresponding to the initial probe beam power, $\bar{\alpha }=({{\alpha }^{+}}+{{\alpha }^{-}})/2$ is the averaged probe-beam absorption coefficient. In the following, we assume $\Delta \alpha ={{\alpha }^{+}}-{{\alpha }^{-}}\ll \bar{\alpha }$ and use the approximation ${{I}_{0}}\exp (-\bar{\alpha }l)\approx {{I}_{T}}$, where ${{I}_{T}}$ is the signal corresponding to the transmitted probe laser power. 

The difference of the refractive index $\Delta n={{n}^{+}}-{{n}^{-}}$ and that of the absorption coefficient $\Delta \alpha ={{\alpha }^{+}}-{{\alpha }^{-}}$ are functions of the probe laser frequency $\omega $ and obey the Kramers-Kronig dispersion relation \cite{Wieman}:
\begin{equation}
\Delta n=-\frac{1}{2{k}_{0}}\Delta \alpha x=-\frac{1}{2{k}_{0}}\Delta {{\alpha }_{0}}\frac{x}{1+{{x}^{2}}},
\end{equation}
where $x=2(\omega -{{\omega }_{0}})/\Gamma$ is a normalized probe detuning from resonant frequency ${{\omega }_{0}}$, $\Gamma $ is the (power- and collision-broadened) homogeneous linewidth (FWHM) of the transition, and the $\Delta {{\alpha }_{0}}$ is the peak value of $\Delta \alpha $ at $x=0$. The signal at the balanced polarimeter is then expressed as
\begin{equation} 
\Delta I=2{{I}_{T}}\Delta \theta =-\frac{{{I}_{T}}}{2}\Delta {{\alpha }_{0}}l\frac{x}{1+{{x}^{2}}},
\label{PSsignal}
\end{equation}
which tells us that the signal obtained by polarization spectroscopy exhibits a dispersion profile with the homogeneous width $\Gamma$ and the maximum amplitude
\begin{equation}
\Delta {{I}_{\max }}=\frac{{I}_{T}}{4}\Delta {{\alpha }_{0}}l. 
\label{Max}
\end{equation}
It is useful to express $\Delta {{\alpha }_{0}}l$ in Eq.(\ref{Max}) as $\Delta {{\alpha }_{0}}l=\Delta \text{OD}^{+}_{\max}-\Delta \text{OD}^{-}_{\max}$, where $\Delta \text{OD}^{+}_{\max}$ ($\Delta \text{OD}^{-}_{\max}$) is the change of the peak OD due to the pump beam for the $\sigma^{+}$ ($\sigma^{-}$) component of the probe beam. The $\sigma^{-}$ component of the probe beam does not interact with the atoms in the $^{1}P_{1}, m_{J}=+1$ state, while the $\sigma^{+}$ component of the probe beam interacts with the atoms both in the  $^{1}P_{1}, m_{J}=+1$ and $^{1}S_{0}, m_{J}=0$ state. If we ignore the effect of velocity-changing collisions, the depth of the dip in the velocity distribution of the atoms in the $^{1}S_{0}, m_{J}=0$ state and the height of the bump in the velocity distribution of the atoms in the $^{1}P_{1}, m_{J}=+1$ state are equal  (see Fig.\ref{Fig5}(b)) and the relation $\Delta \text{OD}^{+}_{\max}=2\Delta \text{OD}^{-}_{\max}$ holds. Equation (\ref{Max}) can then be rewritten as
\begin{equation}
\Delta {{I}_{\max }}=\frac{{I}_{T}}{8}\Delta \text{OD}^{+}_{\max}=\frac{{I}_{T}}{4}\Delta \text{OD}^{-}_{\max}.
\label{Max_OD}
\end{equation}
As long as the approximation ${{I}_{0}}\exp (-\bar{\alpha }l)\approx {{I}_{T}}$ holds, the change of the peak OD due to the pump beam (i.e., the depth of the Lamb-dip in terms of the OD) is given by $\Delta \text{OD}_{\max} = (\Delta \text{OD}^{+}_{\max}+\Delta \text{OD}^{-}_{\max})/2= (3/4)\Delta \text{OD}^{+}_{\max}$. Then, Eq.(\ref{Max_OD}) can also be expressed as  
\begin{equation}
\Delta {{I}_{\max }}=\frac{{I}_{T}}{6}\Delta \text{OD}_{\max}.
\end{equation}
The above expression is quite useful because both ${I}_{T}$ and $\Delta \text{OD}_{\max}$ are easily measured experimentally.

\begin{acknowledgments}
We would like to thank M. Takeuchi, and T. Kuga for helpful discussions. This work was supported by Grant-in-Aid for Challenging Exploratory Research (KAKENHI 23656042) from Japan Society for Promotion of Science (JSPS).
\end{acknowledgments}

\bibliography{Sr_ref}

\end{document}